\documentclass[oldversion]{aa}

\usepackage{graphicx}

\usepackage{txfonts}

\usepackage{natbib}

\begin{document}

\title{A coronal explosion on the flare star CN~Leonis}

\author{J.H.M.M. Schmitt\inst{1}, F. Reale\inst{2}, C. Liefke\inst{1}, U. Wolter\inst{1}, B. Fuhrmeister\inst{1}, 
A. Reiners\inst{3}, G. Peres\inst{2}}

\institute{Hamburger Sternwarte,
Gojenbergsweg 112,D-21029 Hamburg, Germany
\and
Dipartimento di Scienze Fisiche e Astronomiche,
Universit\`a di Palermo, Piazza del Parlamento 1, I-90134 Palermo, Italy
\and
Inst. f. Astrophysik
Universit\"at G\"ottingen, 37077 G\"ottingen, Germany
}


\mail{J.H.M.M. Schmitt, jschmitt@hs.uni-hamburg.de}
\titlerunning{Coronal explosion on CN Leo}
\date{Received / Accepted }

\abstract{
We present simultaneous high-temporal and high-spectral resolution observations at optical and soft X-ray wavelengths
of the nearby flare star  CN Leo.  During our observing campaign a major flare occurred, raising the star's 
instantaneous energy output by almost three orders of magnitude.  The flare shows the often observed {\it impulsive behavior},
with a rapid rise and slow decay in the optical and a broad soft X-ray maximum about 200 seconds after the optical flare peak.
However, in addition to this usually encountered flare phenomenology we find an extremely short ($\tau _{dec}$ $\approx$ 2 sec) soft X-ray peak, which is very likely of thermal, rather than non-thermal nature and temporally coincides with the optical flare peak.  While at hard X-ray energies non-thermal bursts are routinely observed on the Sun at flare onset, thermal soft X-ray bursts on time scales of seconds have never been observed in a solar nor stellar context.   Time-dependent, one-dimensional hydrodynamic modeling of this event
requires an extremely short energy deposition time scale $\tau _{dep}$ of a few seconds to reconcile theory with observations, thus suggesting that we are witnessing the results of a coronal explosion on CN~Leo. Thus 
the flare on CN Leo provides the opportunity to observationally study the physics of the long-sought "micro-flares" thought to be responsible for coronal heating.  
}


\maketitle
\section{Introduction}
The basic energy release in solar and stellar flares is thought to occur in the coronal regions of the underlying
star, by conversion of non-potential magnetic energy through magnetic reconnection \citep{priest02}.  
Particle acceleration takes place and the accelerated particles and/or plasma waves move along the magnetic field lines and reach and heat cooler atmospheric layers \citep{syrat72,brown73}.  Because of their small cooling times 
the heated photospheric layers start radiating immediately, serving as a proxy indicator for
non-thermal particles, while the heated dense chromospheric layers "evaporate" 
leading to a soft X-ray flare \citep{brown73,neupert68,peres00}.
The heating and cooling time scales and the dynamic response of the various atmospheric layers
are important for our understanding of solar and stellar flares and of coronal heating \citep{kostiuk74,somov81,tamres86,fisher87}; a popular hypothesis \citep{klimchuk06,parker88} even attributes all of the required
energy input to the quiescent solar corona to (nano)-flare input.  

Solar and stellar flares are observed to occur over vastly different time scales ranging from a few seconds \citep{vilmer94,schmitt93} to more than a week \citep{kuerster96}, and the response of the heated plasma sensitively depends on the mode of energy input. 
Since the flare process involves plasma differing in temperature and density by more than four orders of magnitude, 
varying on short spatial and temporal scales, simultaneous multiwavelength data with sufficient spectral coverage and 
sufficient spectral and temporal resolution are required for any sensible observational diagnostics.  Since the primary energy release and energy loss process of a flare is coronal, space-based X-ray observations are required. 
Using ESA's XMM-Newton X-ray observatory and ESO's Ultraviolet-Visual Echelle Spectrograph (UVES) 
on Kueyen on May 19$^{th}$ 2006  we observed a spectacular flare on the nearby (d~=~2.39~pc) flare star CN~Leo (spectral type: M5.5-M6.0, $T_{eff}$ = 2800-K~2900~K).  Despite its low
apparent rotational velocity of $v\ \sin i < 3.0$~km~s$^{-1}$ \citep{fuhr04}, CN~Leo shows all the attributes of
a magnetically active star, including chromospheric and coronal emission as well as flaring in the
optical and in the X-ray bands \citep{fuhr07}.  

\section{Observations and data reduction}

The \emph{XMM-Newton} observatory carries three co-aligned X-ray telescopes and a co-aligned smaller optical telescope 
equipped with the Optical Monitor (OM); this satellite and its instruments are described in detail
in a Special Issue of Astronomy \& Astrophysics Vol. 365 (Jan. 2001).  For the CN~Leo observations we used the medium filter 
in Full Frame and Large Window mode for all EPIC observations.
The OM was operated in fast window mode, allowing U band flux measurements
with a time resolution of 1 sec, limited by the chosen read-out sequence.
The X-ray data were reduced with the \emph{XMM-Newton} Science Analysis System 
(SAS) software, version 7.0.  The basic \emph{XMM-Newton} data consist of 
individual photons with known arrival time, position and energy; using standard filtering 
criteria these data are further processed into EPIC lightcurves and spectra. 

\begin{figure*}
\includegraphics[width=17.0cm]{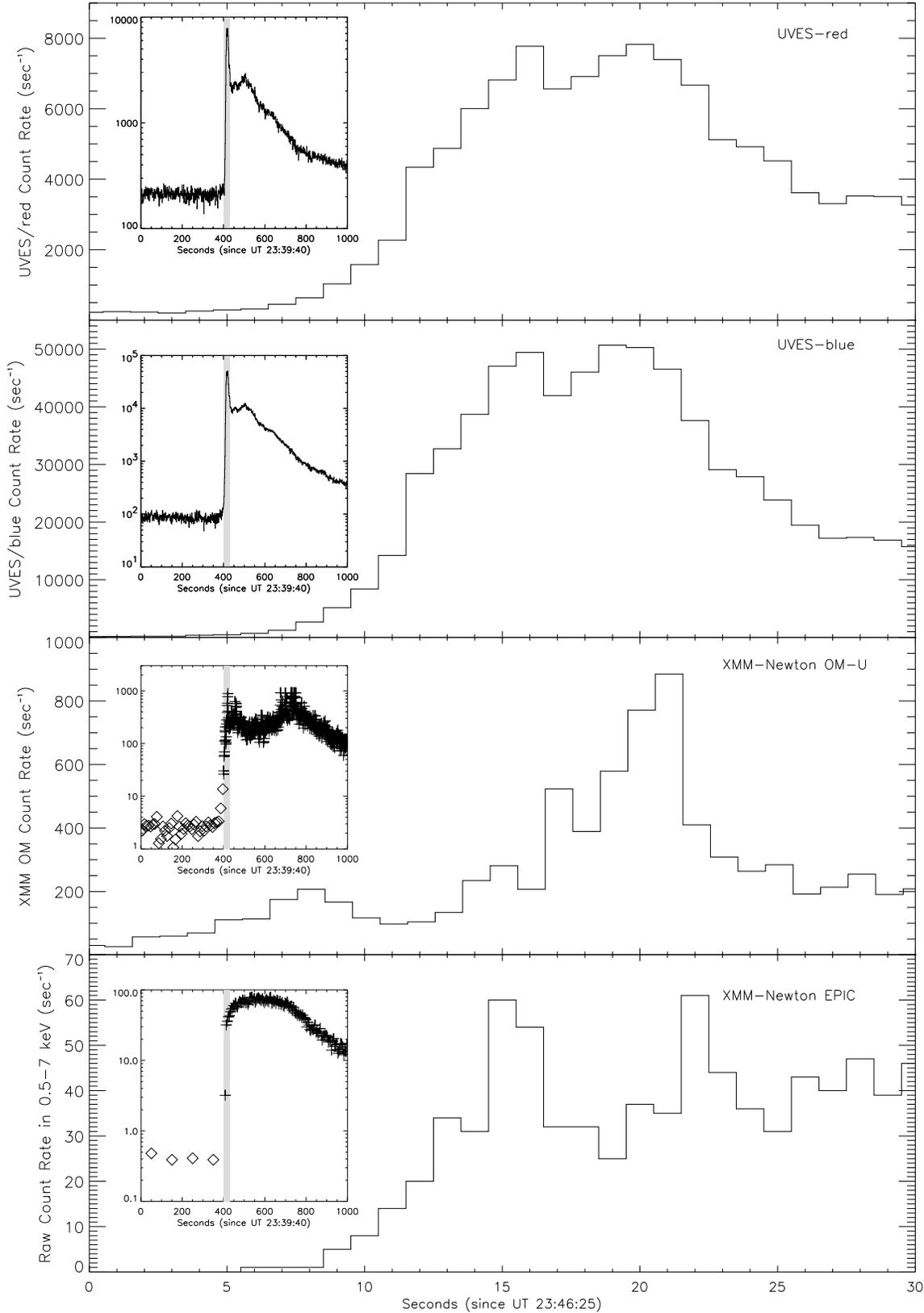}
\caption{\label{light1} XMM-Newton EPIC raw X-ray (sum of PN,MOS1,MOS2 detectors)
and OM U band light curves (bottom panels),
UVES blue and red channel (upper panels) during the first 30 seconds of the impulsive  flare on CN~Leo ; all times are in UT seconds 
relative to May, 19$^{th}$, 2006, 23:46:25;  the inserts show the same light curves covering 1000 seconds relative to May, 19$^{th}$, 
2006, 23:39:40 in logarithmic scale including some quiescent emission and the full flare light curve; the light curve portions displayed
enlarged are gray shaded.
The X-ray light curves exclude events recorded in the center of the point response function to avoid pile-up effects; the total X-ray count rate is almost twice as large. The binning of all data is 1 sec, except that of the X-ray data in the insert, which is set to 100 sec prior 
to the flare onset and to 10 sec afterwards.
The OM light curve is dead time corrected; for rates above 500 cts/sec these corrections and the lightcurve
become unreliable.}
\end{figure*}

UVES is an cross-dispersed echelle spectrograph mounted on the Nasmyth~B
focus of the Kueyen telescope (UT2) covering the wavelength range from 
about 3000 \AA\ to 11\,000~\AA\  with a typical spectral resolution  
of $\sim$ 40000.  The temporal resolution of the high-resolution spectra is
determined by the sum of exposure and read-out times (a few minutes in our case),
a small fraction of the optical light is directed into an exposuremeter for each of the two
spectral bands ("blue" and "red"), thus providing broad-band photometry with a time resolution of one second.
These data are, however, taken mostly for engineering purposes, and are not corrected for background or airmass.  
For our analysis we used the UVES pipeline products with the 
wavelength calibration carried out using Thorium-Argon spectra
resulting in an accuracy of $\sim 0.03$~\AA\, in the blue arm and 
$\sim 0.05$~\AA\, in the red arm. Absolute flux calibration is based on
the UVES master response curve and extinction files provided by ESO. 

\section{Results}

\subsection{X-ray and optical light curves}

In Fig.~\ref{light1} we plot -- from bottom to top -- the recorded XMM-Newton X-ray light curve (i.e., the sum of the PN+MOS1+MOS2 detectors), 
the XMM-Newton Optical Monitor (OM) U~band light curve, and the UVES blue and red light curves for part of our observing run. 
In each panel we show a linear light curve covering 30 seconds around the flare onset, while the respective inserts show the logarithmic light curves of the
whole flare event including some quiescence.  The most notable feature in these light curves is the enormous, simultaneously occurring
intensity increase in all considered energy bands starting
at around UT = 23:46:20.  The relative increase in terms of flare peak to pre-flare flux is $>$ 100 in X-rays and $\approx$ 750 in
the UVES "blue" band.  Even in flare stars flares with magnitude increases $>$ 7
as shown in Fig.~\ref{light1} are quite rare, and we are actually not aware of any examples in the literature
sampled with the time resolution and spectral coverage of our data.
The temporal alignment of the light curves in different bands ought to be accurate to at least 0.5 seconds and hence below the bin width.
Fig.~\ref{light1} shows four remarkable properties of our multiwavelength light curves: First, the flare onset appears to be earlier in the optical wavebands, in particular, there is no evidence for any significant soft X-ray "precursor" activity, while there is some "pre-flash" visible in the U band; 
second, the rise from quiescence to flare peak in the optical is 750~fold and occurs within 
$\approx$ 10 seconds; third, both the "red" and "blue" UVES light curves show two peaks
separated by 5 seconds; and fourth, the first optical peak is accompanied by a soft X-ray peak.   Note that this
first soft X-ray peak is very sharp, lasting at most  $ \sim $ 2 seconds, and does not coincide with 
the main soft X-ray flare peak, which in fact is quite flat and occurs about 200 seconds later (cf., insert in bottom panel of Fig.\ref{light1}). 

\begin{figure}[h]
\includegraphics[width=7.0cm,angle=-90.]{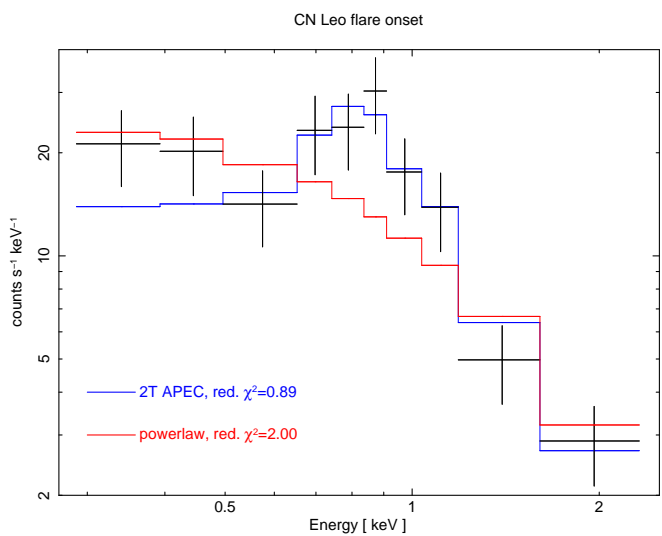}
\caption{\label{onset} EPIC pn spectrum of burst between UT~23:46:36 to  UT~23:46:44 (black data points) together with model fit assuming a thermal
spectrum (blue curve) and a power law (red curve).
}
\end{figure}

\begin{figure}[h]
\includegraphics[width=7.0cm,angle=-90.]{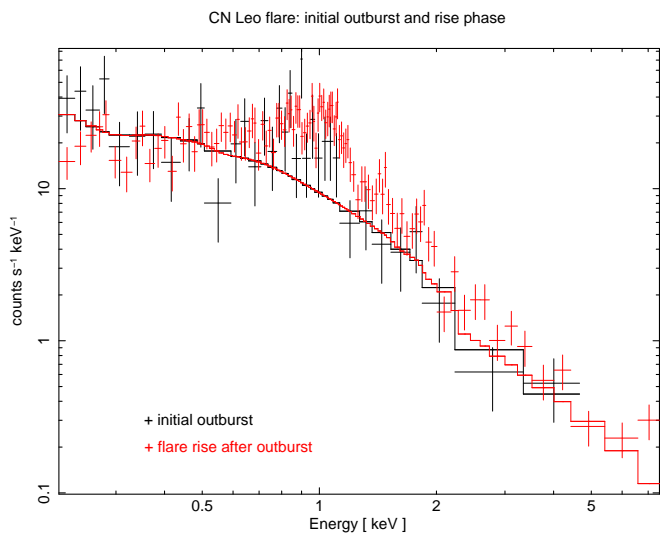}
\caption{\label{onset2} Comparison of burst spectrum between UT~23:46:36 to  UT~23:46:44 (black data points) with 
postburst spectrum (red data points) between UT~23:46:44 to  UT~23:47:44; a power law fit with photon index $\gamma = 2$ is also shown.
}
\end{figure}
\subsection{The spectral nature of the X-ray burst}

The crucial question concerning the X-ray light curve is the nature of the soft X-ray peak at time T = 15 sec in Fig.\ref{light1}. 
We emphasize that this peak is statistically highly significant. We find a total of 105 counts in the two adjacent 1~sec bins 
around the optical flare maximum.   A smooth interpolation of the flare rise light curve leads us to a "background" 
estimate of $\approx$ 55 counts in those two bins, implying an extremely high statistical significance of the 2 second flare peak.  
Accepting then the X-ray and optical peak and their close temporal association as real,
on the one hand would suggest a non-thermal origin of the observed X-ray emission, while on the other hand, non-thermal
X-ray emission has never been conclusively demonstrated from stars \citep{osten07}. Also, impulsive {\bf soft} X-ray radiation is
known from the Sun only on much longer time scales \citep{hudson94}. For the Sun, non-thermal, often impulsive X-ray emission is 
routinely observed, but below $\sim$ 25 keV the discrimination between thermal and non-thermal emission becomes exceedingly difficult \citep{kahler75}.  On the other hand,
assuming a thermal origin of the first X-ray peak makes the observed fast decay time of $\sim$ 2 sec difficult to reconcile with the usually invoked conductive or radiative cooling time scales.  

Unfortunately the spectral analysis of the EPIC pulse height data in the respective time interval from 
UT~23:46:36 to  UT~23:46:44 is plagued by the low SNR in this short stretch of data.
The rebinned EPIC-pn spectrum of this time interval is shown in Fig. \ref{onset}, together with a best fit assuming a thermal (blue curve) and a power law (red curve) spectral shape.  For the thermal fit we assumed a two-temperature model with solar abundance, incidentally,
a one-temperature component model with variable abundance also yields fits of comparable statistical quality.
The fit quality of a power law fit ($\chi^2_{red} = 2.0$) is clearly far inferior to that of the thermal models ($\chi^2_{red} = 0.9$) , 
and a an improvement in quality of a power law fit can only be brought about by the introduction of substantial amounts of cold absorption ($N_H \sim 4 \times 10^{21}$ cm$^{-2}$), which we regard as utterly unphysical in the context of CN~Leo.  Therefore we clearly prefer a thermal interpretation of the "burst" spectrum purely on observational grounds.  This thermal interpretation is supported by the
X-ray spectrum between UT~23:46:44 to UT~23:47:44, immediately after the burst,
displayed in Fig.·\ref{onset2}, together with the power fit producing an acceptable fit to the burst spectrum
in Fig.\ref{onset}.  Clearly, this spectrum cannot possibly be fit with a power law spectrum, the iron L complex and Ne lines in the
energy range 0.7-1.2 keV prevent a power law description; formally, a power law fit with variable slope
results in ($\chi^2_{red} = 3.32$), while a thermal spectrum yields ($\chi^2_{red} = 1.01$) and thus we conclude that the spectrum between  between UT~23:46:44 to  UT~23:47:44 is definitely thermal.  
Going back to Fig. \ref{onset} we realize that the largest power law fit residuals are found near 1 keV, precisely in that spectral range where the thermal nature of the post-burst spectrum becomes most apparent.  If we assumed the burst spectrum to be non-thermal,
the best fit power law slope of $\gamma = 2$ of the burst spectrum (cf., Fig.\ref{onset}) would have to agree only coincidentally
with the "slope" of the thermal spectrum in the post burst spectrum (cf., Fig.\ref{onset2}).
Therefore we conclude that the EPIC pulse height spectra provide no proof for a non-thermal character of the
observed X-ray emission during any time in the flare.

\begin{figure}[h]
\includegraphics[width=9.0cm,height=5cm]{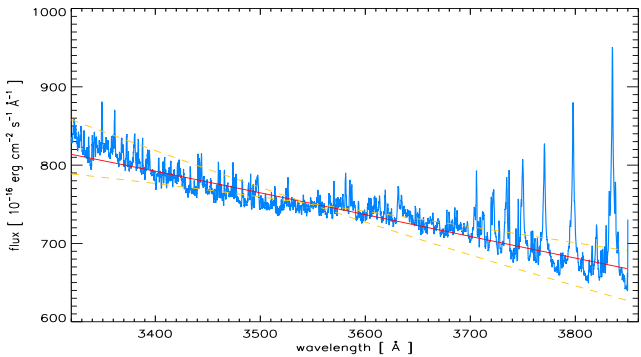}
\caption{\label{spec} VLT/UVES spectrum (between UT~23:30:35 to 23:47:15) covering the wavelength range 
between 3300 - 3850 \AA\ .  Data points are shown as blue histogram, black body fits with temperatures
of 11318~K in red and for 10~000~K and 15~000~K as dashed lines.  Balmer line emission
is visible at longer wavelengths, but at shorter wavelengths most of the flux resides in continuum.
}
\end{figure}
\subsection{Optical burst spectrum}

A high-resolution UVES spectrum covering the time interval UT~23:30:35 to UT~23:47:15 covering the wavelength range 
between 3200 \AA\ - 3850 \AA\ has been taken (see Fig.\ref{spec}). 
While this spectrum covers also periods of quiescence, it is clearly dominated by the flare and can be taken
as a representative mean spectrum of the first minute of flare emission.   
Over this wavelength range the emission is
almost linearly decreasing with wavelength without any major emission lines except the Balmer lines at the long wavelength end; the bulk of the chromospheric emission
lines appears only in later spectra.   The slope of the best fit blackbody spectrum spectrum corresponds to an effective temperature
of $\sim$ 11~320~K; other fits with temperatures fixed at 10~000~K and 15~000~K are also shown. 
Converting then the observed spectral normalization to an emission area (using the known distance towards
CN~Leo) results in an area estimate of $\approx$ 4.4 $\times$ 10$^{18}$~cm$^2$, while the other temperatures lead to values
between 1.7 - 7.1 $\times$ 10$^{18}$~cm$^2$.
Some caution is warranted in this estimate:  As
is evident from Fig.\ref{light1}, significant photospheric temperature changes take place during the first 60 seconds, which
cannot be resolved in the UVES spectrum.  The derived temperature is therefore a mean and probably lower than
the achieved maximum temperature.  Further, no calibration spectra of standard stars were taken during the same night and using
the spectra of different standard stars taken during different nights leads to errors in the absolute calibration 
of up to 50 \%.  We therefore conclude that the emission area responsible for the observed blue optical emission was
in the range 1-10 $\times$ 10$^{18}$ cm$^2$. An independent estimate of the size of this emission
area can be derived from the UVES photometer data.
Using the mean temperature derived from the spectrum and the count rate during the spectral exposure, we can
convert the observed count rate into an energy flux (from a comparison with other stars) and derive in this fashion
an emission area of $\approx$ 8$\times$ 10$^{18}$ cm$^2$, in good agreement with the estimate from the spectrum.  We thus
conclude that the photospheric flare peak temperature was in excess of 11~000 K and covered a flare area $A_{fl}$
of 1-10 $\times$ 10$^{18}$ cm$^2$ and obviously one expects the X-ray emission to arise from a similarly-sized region.
Using the pile-up corrected PN light curve we can use the observed flare count rate 
in the energy band 0.5 - 8 keV to compute an (isotropic) radiative loss of $\approx$ 3.8 10$^{31}$ erg; we estimate that
about two thirds of the overall radiative losses are contained in this energy band.  During the first X-ray peak, the recorded
X-ray luminosity is estimated to be $\sim$ 5 $\times$ 10$^{28}$ erg s$^{-1}$ (Fig. \ref{light3}).


\begin{figure}[h]
\includegraphics[width=9.5cm,height=12cm]{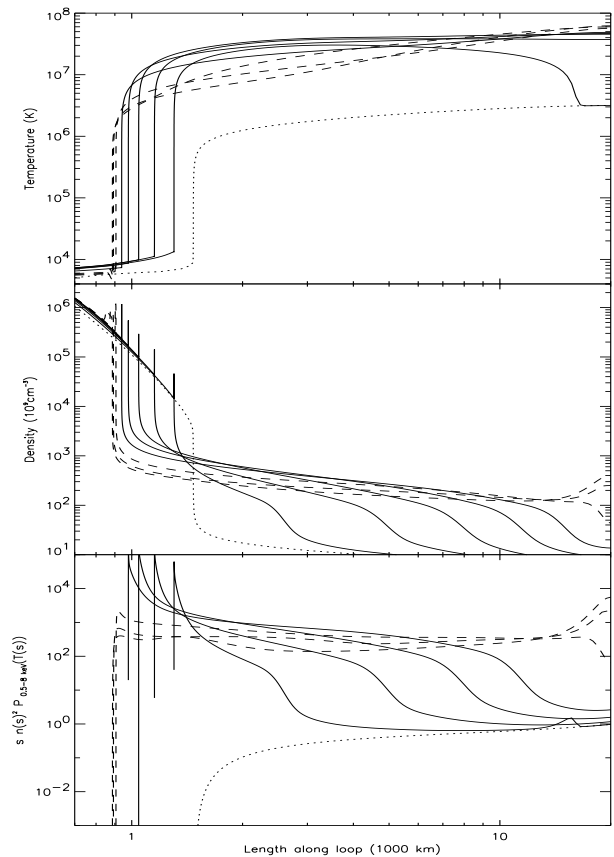}
\caption{\label{simul1}
Calculated run of temperature (upper panel), density (in units of 10$^9$ cm$^{-3}$; medium panel) and emission measure weighted
cooling function (bottom panel)
for the first 20 seconds of a coronal
explosion.   The initial state (a quiescent loop with T$_{max} = 3 \times 10^6$ K) is shown as dotted line, the evolution at times
t = 2s, 4s, 8s and 10s, during which heating takes place,  is shown as solid lines, that at times t = 13s, 16s, 
and 19s as long-dashed lines.}
\end{figure}

\section{Modeling and Interpretation}
\subsection{Model ansatz}
To address the question whether our observations and the above derived physical parameters
are consistent with the expected hydrodynamic response of
coronal plasma to a fast transient heat flux we performed a
hydrodynamical simulation of an initially hydrostatic loop with prescribed length $L=4 \times 10^9$ cm and apex 
temperature $T_{max} = 2 \times 10^6$ K subjected to impulsive heating.
For our hydrodynamical simulations we use the Palermo-Harvard (PH) code \citep{peres82,betta97},
which solves the time-dependent 1-D hydrodynamic equations on an Eulerian grid.  
The energy equation includes the effects of optically thin radiative losses, heating and
thermal conduction, treated in the Spitzer form, which should apply in a high-density medium, where the heat pulse is deposited,
The heating term Q consists of two contributions, a spatially and temporally constant term
to produce a quiescent corona,
and a spatially and temporally varying term to describe a flare.
Since the microphysics of the heating is not well understood, we use
an {\it ad hoc} parameterization of the form $Q_{fl} = H_0 f(t) g(s)$.
For $g(s)$ we use a Gaussian $g(s) = \frac{1} {\sqrt{(2 \pi \sigma)^2}} exp(\frac{-(s-s_0)^2}{2 \sigma ^2})$,
centered at $s_0 = 2 \times 10^8$ cm (as measured from the loop footpoint)
and $\sigma = 10^8$ cm, for f(t) we
use a symmetric triangular profile such that f(t) differs from 0 for a time interval
$t_{heat}$ and assumes a maximal value of unity and $H_0$ =  7200 erg cm$^{-3}$ s$^{-1}$.  The main difficulty
for a numerical solution of the corona problem is the treatment of the
transition region, where temperature and density gradients are extremely large.
These numerical difficulties are aggravated by the fact that during the
flare evolution the position of the transition region changes its spatial
location.  The PH code uses a re-adaptive grid, which insures a maximal
relative change of 10 \% in all physical variables on adjacent grid
points, provides a smoothly varying grid size over the whole
computational domain and samples the whole region of interest with
typically 500 - 550 grid points. At the footpoints the model loop is anchored to a dense 
chromosphere, the structure of which follows the treatment by \citet{vernazza81} and
in which the energy balance is maintained throughout the simulation.
We assume the loop to be symmetric with respect to its apex, so that the evolution
is computed along one loop half and symmetry boundary conditions are
imposed at the apex. The lower boundary is maintained at a fixed
temperature at the chromospheric temperature minimum.

\subsection{Explosion modeling}

We specifically consider a simulation with $\tau_{dep}$ = 10 sec and a total deposited heat fluence
E$_{tot}$ = 1.8 10$^{13}$ erg/cm$^2$, assumed to be deposited at a height of 2000 km above the photosphere.  
The results of our simulation are shown in Fig. \ref{simul1}, where we plot the run of density and temperature vs. loop length for the initial state and every 2 seconds into the flare evolution.
The main effect of the transient heating is a rapid temperature increase
of the "old" (i.e., pre-flare) corona and transition region of the loop
and parts of the "old" chromosphere to temperatures of 20 - 40 MK.
Because of the high-density of the latter, a high-pressure region,
sandwiched between the low-pressure "old" corona and low-temperature
unheated chromosphere/photosphere,
is formed at the bottom of the "old" transition region.  An expansion
wave starts propagating upwards with speeds in excess of 1000 km/s
to increase the density of the "old" corona ("loop filling").
This explosive filling of the loop occurs on a time scale of 10 seconds, comparable to the sound
crossing time of the structure.  Once the loop is rapidly filled, the material is more gently accumulated and compressionally
heated, leading to a slower rise in emission measure and temperature.
At the same time, the high-pressure "new" transition region moves inwards
into the chromosphere with speeds of $\sim$ 30 km/s; a narrow
density spike separates inward moving outer regions from not yet accelerated inner regions.  
Once the heating stops, the temperature in the lower loop
portions drops through conductive losses into the chromosphere,
while the temperature at the loop top still increases through
compressional heating.  From the runs of temperature and density we
compute the temperature-dependent emission measure and 
cooling in the energy range 0.5-8 keV, displayed in Fig. \ref{simul1}.  Obviously during the first $\approx$~10~seconds the main contribution
to the X-ray flux comes from the bottom parts of the loop, while at later times the evaporated plasma dominates
the overall coronal emission.

\begin{figure}[h]
\includegraphics[height=8cm]{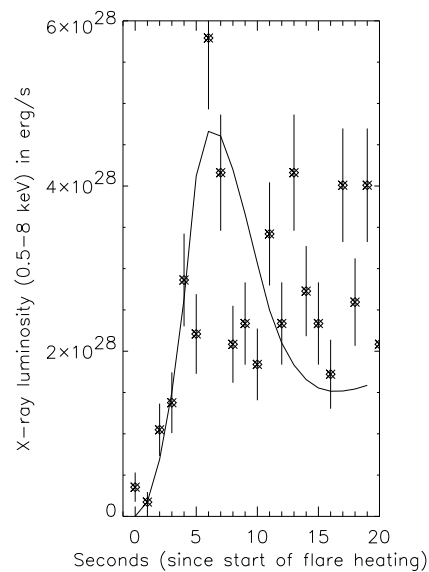}
\caption{\label{light3} XMM-Newton X-ray light curve (in erg/s)
overplotted over hydrodynamic model prediction; time is with respect to
the modeled start of the energy input.
Note the good agreement between data and model although no best fit was performed.  Also note that the event at 
t = 13 sec might represent a further soft X-ray burst.
}
\end{figure}

\subsection{Comparison to observations and interpretation}

The theoretical curves displayed in Fig.\ref{simul1} can be readily compared to our XMM-Newton data.
It is straightforward to compute the total
X-ray flux assuming some loop cross sectional area and in Fig. \ref{light3} we 
juxtapose our model predictions (solid line) computed under the assumption of an emission area of
$A_{fl}$ = 10$^{19}$ cm$^2$ and the observed X-ray luminosities computed from the pile-up corrected PN lightcurve
using a count-rate-to-flux conversion of 1.8 $\times$ 10$^{-12}$ erg/cm$^2$/count and the known distance towards CN~Leo.
Although no best fit (in terms of the temporal model evolution) has been attempted we note the good agreement between
theoretical expectations and observations. The X-ray peak coincides
with the end of the heating phase and the spike in the light curve
at t = 13 sec may represent another soft X-ray burst.

\begin{figure}[h]
\includegraphics[height=7cm]{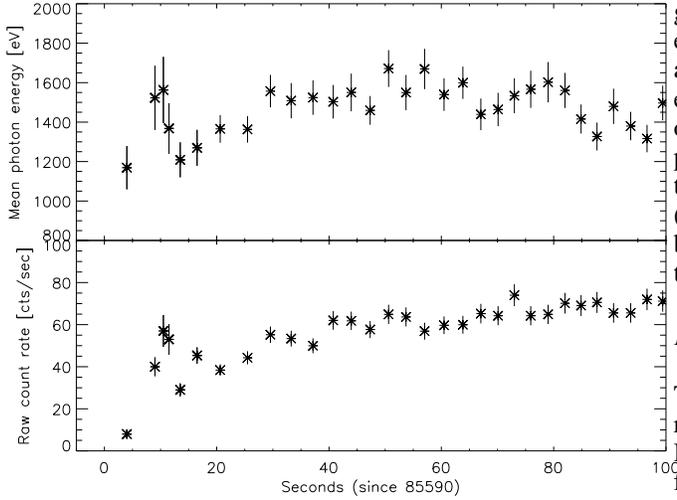}
\caption{\label{light4}  XMM-Newton EPIC light curve (bottom panel) and mean photon energy (top panel) vs. time
for the first 100 seconds around flare onset; all times are relative to  May, 19$^{th}$, 2006, 23:46:20.  The mean photon
energy has been derived from all photons with registered energies $>$ 500 eV; each data point represents the mean
energy of 200 photons.
}
\end{figure}

\subsection{Comparison to observations and interpretation}

The above interpretation of the XMM-Newton light curve is supported by the observed
run of the evolution of count rate and mean photon energy for the first
100 seconds into the flare shown in Fig.\ref{light4}.
As is clear, the mean photon energy and therefore the mean X-ray temperature are decreasing, implying
that the X-ray emitting plasma is cooling. In the downward direction conductive cooling is the dominant cooling process;
computing the cooling time scale at the heat pulse center through conduction via
$ \tau_{cond} = \frac {21 n k} {2 \kappa _0} \frac {L^2} {T^{5/2}} $ we find $\tau_{cond} \approx $ 1.7 sec using 
n = 7 $\times$ 10$^{12}$ cm$^{-3}$, L = 1 $\times$ 10$^{8}$ cm and T = 3.4 $\times$ 10$^{7}$ K.
The rapid cooling in the lower section of the loop leads to a pressure deficit accompanied by a density 
decrease and ensuing decrease in X-ray emission (cf., Fig.\ref{simul1}),
driven by the fast evaporation front moving ahead. 
The short lasting soft X-ray burst shown in Fig.\ref{light1}
is thus explained with a highly coherent plasma evolution,
i.e., by flaring plasma confined in a single magnetic flux tube and
by assuming that a major fraction or all of the observed energy release of $>$ 3.8 10$^{31}$ erg occurred on a rather short time scale
of 10 seconds or less.  We therefore conclude that we are directly witnessing the process of evaporation as a result of the explosive
heating of the lower chromosphere. 
 
\subsection{Thick target bremsstrahlung}

As is clear from the above description the energy input in our modeling occurs 
in an {\it ad hoc} fashion by specifying an arbitrary spatially and temporally varying heating function.  
Since the precise nature of the heating physics is unknown, this approach seems justified.  Of course, the physical 
picture we have in the back of our minds is that of energetic particles
(electrons) penetrating deep into the chromosphere and upper photosphere, dissipating their energy there and heating up the adjacent atmosphere.  
Since the UVES light curves are well co-aligned with the initial X-ray light curve and since the former emission
comes from relatively cool material far extending into the red wavelength band, we require fairly energetic electrons
to energize these photospheric layers.
Unfortunately, the precise properties of these electrons and, in particular, their energy spectra and spectral cutoffs
are only poorly known for solar flares and essentially unknown in the case of stellar flares.  At any rate, one expects thick-target bremsstrahlung from this non-thermal particle population, but at what level ? To this end we follow the calculations by
\citet{syrat72} and \citet{brown73}, who considered the bremsstrahlung X-ray spectra for bremsstrahlung emitted by a population of electrons injected into the flare region with an energy distribution function of the form 
\begin{equation}
F(E) = F_0 (\frac{E_0}{E})^{\gamma}.
\label{eq1}
\end{equation}
The approach taken by \citet{brown73} is analytical, the most recent calculations of thick-target 
bremsstrahlung are those by \citet{holman02}, who computed microwave and X-ray spectra for such power law electron
distributions using the appropriate relativistic cross sections and considering the effects of low and high energy cutoffs.   
As demonstrated by \citet{brown73} and \citet{syrat72}, an electron power law distribution of the form given in eq.\ref{eq1} leads to
a power law distribution of the observed photon flux density $I(\epsilon)$ of the form
\begin{equation}
I(\epsilon) = I_0 (\frac{E_0}{\epsilon})^{\beta},
\label{eq2}
\end{equation}
i.e., again a power law distribution.  The respective power law indices and normalizations are related through
\begin{equation}
\beta = \gamma - 1
\hfil\break
F_0 E_0= \frac{\beta (\beta -1) I_0}{A B(\beta-1,1/2)};
\label{eq3}
\end{equation}
$B(\beta-1,1/2)$ in eq. \ref{eq3} denotes the Beta-function, and the constant A is given by
\begin{equation}
A = \frac {8}{3} \frac {r_o ^2}{137} \frac {m c^2} {\pi e^4 Z^2 \Lambda},
\label{eq4}
\end{equation}
where the quantities $c$,$r_o$,$m$,$e$,$Z$ and $\Lambda$ denote speed of light, classical electron radius, electron mass,
electron charge, ion charge number, and Coulomb logarithm respectively.  Finally, the total energy content $E_{tot}$ of the electron
population in the energy range from $E_0$ to infinity can be computed from
\begin{equation}
E_{ŧot} = \frac {I_0 \beta E_0} {A B(\beta -1, 1/2)}
\label{eq5}
\end{equation}
 Thus, as well known from solar physics,  the power law index of the electron distribution as well as the normalization
of the electron distribution and therefore its total energy content can be inferred from the observed photon power law index $\beta$
and the observed photon fluence $I_0$,
{\bf if} the photons are indeed generated by the thick target bremsstrahlung mechanism.  If we assume -- for argument's sake --
a thick target bremsstrahlung origin
for the observed burst, we use the best-fit photon spectral index of $\beta \sim 2$ and the burst integrated photon fluence
at 1 keV of 4.8 $\times$ 10$^{37}$ photons/keV to compute the total energy content $E_{pl}$ of the power law electron population of 
2.4 $\times 10^{35}$ ergs with a low energy cutoff of 1 keV; assuming a low energy cutoff of 0.1 keV, we instead find 
$E_{pl} = 2.4 \times 10^{36}$ erg.  Regardless of the assumed low energy cutoff, the
thus derived total energies are a few orders of magnitude above the observed soft X-ray energy output of 3.8 $\times 10^{31}$ erg.
In other words, under the assumption of non-thermal bremsstrahlung we would have to assume that unlike the solar case the bulk of the energy contained in the non-thermal electron population goes into other (unobservable) forms of energy.  Applying Occam's razor we 
thus conclude that both observations and theory suggest that the observed soft X-ray emission is {\bf not} due to thick-target bremsstrahlung.  At the same time we have shown that thick-target bremsstrahlung, which is of course expected to be produced at some level, can be easily accommodated with in the existing observational framework, even if its low energy cutoff were to extend down to below 1 keV and therefore our theoretical modeling is fully consistent with the existing data.

\subsection{Physical consistency of modeling}

Does our theoretical modeling capture all of the important physics of the flare explosion~?
A basic assumption in our modeling is the 1-D character of the problem.  Enormous pressure differences exist between
"bottom" and "top" of the flaring plasma, while the magnetic field imposes a strong 1-dimensional directionality
of the plasma along the field lines.   Inspecting
the maximally occurring gas pressures, we find values of up to
10000 dyn cm$^{-2}$, requiring a magnetic field of $>$ 300~G for confinement.   The mean photospheric field in CN~Leo
was simultaneously measured to exceed 2000 G \citep{reiners07},
thus plasma confinement is no problem and our 1-D modeling should be correct.
The released energy is assumed to come from magnetic reconnection.  An efficient energy conversion requires the incoming 
flow velocity to be less than 10~\% of the Alfven speed v$_A$ and hence the time scale $\tau_{diss}$ for an efficient magnetic field
dissipation on a spatial scale $L$ becomes $\tau_{diss} \sim \frac{L} {0.1 v_A}$.  
Using $L \sim 2 \times 10^9$ cm and $\tau_{diss} \sim $~5~seconds, we find v$_A \sim 4 \times 10^9$ cm/s, which appears
unrealistically large.  However, since we expect coronal fields on the order of $\approx$ 300 G and particle densities 
of $\approx$ 5 10$^{8}$ cm$^{-3}$, we find Alfven velocities of v$_A \sim 3 \times 10^9$ cm/s, and therefore the assumption of
Alfven speeds in excess to $10^9$ cm/s is reasonable.  

\section{Discussion and conclusions}

Heating by so-called "microflares" and "nanoflares" \citep{cargill04,klimchuk06} is a popular hypothesis to explain the heating of the corona of the Sun and that of other stars.  Their nanoflare model assumes that any coronal loop is composed of many unresolved strands of magnetic flux and that these
strands  are "heated impulsively by a small burst of energy (which, for convenience, are called a nanoflare, although the range of energies is completely arbitrary". 
Further, the strands are thermally isolated from each other and can be viewed as elemental magnetic flux tubes with a small value for the plasma beta
\citep{cargill04}. An individual heating event on such a strand leads to a dynamic evolution of the affected plasma
located on this particular strand.  In a typical observation of a star and even in a typical observation of 
the solar corona, a multitude of heating events occur more or less simultaneously on different strands located in the field of view or are integrated
over during a typical exposure time. Therefore
in this picture any solar or stellar coronal observation represents a mean of the heating and cooling history of possibly a
very large number of individual energy releases.  For a computation of the global properties of solar and stellar coronae in terms of the distributions
of plasma densities, temperatures and flow velocities, a knowledge and modeling of the relevant heating and cooling processes
is mandatory. Such models specifically assume an impulsive heating of the strands of magnetic flux, which can be recognized only in data with
very high time resolution ($\sim$ 1 sec), and then cool, first, by conduction
and then by radiation \citep{cargill04}.  This is precisely what appears to have happened on CN~Leo:  An energy release  
on a time scale of less than 10 sec in a single loop has to our knowledge never been observed in any star nor on the Sun. 
Therefore, with our high time-resolution soft X-ray and optical observations of the impulsive flare on CN~Leo -- by its energetics definitely not a 
nanoflare -- we may have isolated the relevant physics for nanoflare heating of the corona of the Sun and that of the stars. 

\begin{acknowledgements}
This work is based on observations obtained with {\em XMM}-Newton,
an ESA science mission with instruments and contributions directly
funded by ESA Member States and the USA (NASA). and on observations 
collected at ESO, Paranal, Chile, under the program 076.D-0024(A).
C.L., U.W. and B.F. acknowledge financial support by the DLR under 50OR0105,
AR through an Emmy Noether Fellowship from DFG under DFG RE 1664/4-1.
\end{acknowledgements}


\begin{thebibliography}{}

\bibitem[\protect\astroncite{Betta et al.}{1997}]{betta97}
Betta, B., Peres, G., Reale, F., Serio, S., 1997, A\&AS, 122, 585

\bibitem[\protect\astroncite{Brown}{1973}]{brown73}
Brown, J.C., 1973, Sol. Phys., 28, 151-427

\bibitem[\protect\astroncite{Cargill and Klimchuk}{2004}]{cargill04}
Cargill, P.J., Klimchuk, J.A., 2004, ApJ, 605, 911-920

\bibitem[\protect\astroncite{Fisher}{1987}]{fisher87}
Fisher, G.H., 1987, ApJ, 317, 502-513

\bibitem[\protect\astroncite{Fuhrmeister et al.}{2004}]{fuhr04}
Fuhrmeister, B., Schmitt, J.H.M.M., Wichmann, R., 2004, A\&A, 417, 701-713, (2004)

\bibitem[\protect\astroncite{Fuhrmeister et al.}{2007}]{fuhr07}
Fuhrmeister, B., Liefke, C., Schmitt, J.H.M.M., 2007, A\&A, 468, 221- 231


\bibitem[\protect\astroncite{Holman}{2002}]{holman02}
Holman, G., 2002, ApJ, 586, 606-616 

\bibitem[\protect\astroncite{Hudson et al.}{1994}]{hudson94}
Hudson, H.S., Strong, K.T., Dennis, B.R., Zarro, D., Inda, M., Kosugi, T., \& Sakao, T., 1994, ApJ, 422, L25-L27

\bibitem[\protect\astroncite{Kahler}{1975}]{kahler75}
Kahler, S., 1975, in: Solar gamma-, X-, and EUV radiation; Proceedings of the Symposium, Buenos Aires, Argentina, June 11-14, 1974. (A76-10126 01-92) Dordrecht, D. Reidel Publishing Co., 211-231

\bibitem[\protect\astroncite{Kostiuk and Pikelner}{1974}]{kostiuk74}
Kostiuk, N.D., Pikelner, S.B., 1974, Astron. Zh., 51, 1002-1016

\bibitem[\protect\astroncite{Klimchuk}{2006}]{klimchuk06}
Klimchuk, J., 2006, Solar Physics, 234, 41-77

\bibitem[\protect\astroncite{K\"urster and Schmitt}{1996}]{kuerster96}
K\"urster, M., Schmitt, J.H.M.M., 1996, A\&A, 311, 211-229


\bibitem[\protect\astroncite{Neupert}{1978}]{neupert68}
Neupert, W.M., 1968, ApJ, 153, L59-62

\bibitem[\protect\astroncite{Osten et al.}{2007}]{osten07}
Osten,R.~A., Drake, S., Tueller, J., Cummings, J., Perri, M., Moretti, A. and Covino, S.,2007, ApJ, 654, 1022-1067

\bibitem[\protect\astroncite{Parker}{1988}]{parker88}
Parker, E.~N., 1988, ApJ, 330, 474-479

\bibitem[\protect\astroncite{Peres et al.}{1982}]{peres82}
Peres, G., Rosner, R., Serio, S., Vaiana, G.S., 1982, ApJ, 252, 791-799

\bibitem[\protect\astroncite{Peres}{2000}]{peres00}
Peres, G., 2000, Solar Physics, 193, 33-52

\bibitem[\protect\astroncite{Priest and Forbes}{2002}]{priest02}
Priest, E.R., Forbes, T.G., 2002, A\&ARev, 10, 313-377

\bibitem[\protect\astroncite{Reiners et al.}{2007}]{reiners07}
Reiners, A., Schmitt, J.H.M.M., Liefke, C., 2007, A\&A, 466, L13-L16, (2007)

\bibitem[\protect\astroncite{Schmitt et al.}{1993}]{schmitt93}
Schmitt, J.H.M.M., Haisch, B.M., Barwig, H., 1993, ApJ, 419, L81-84

\bibitem[\protect\astroncite{Somov et al.}{1981}]{somov81}
Somov, B.V., Syratovskii, S.I., Spektor, A.R., 1981, {\it Solar Physics}, 73, 145-155

\bibitem[\protect\astroncite{Somov et al.}{1997}]{somov97}
Somov,B.V., Kosugi, T., Litvinenko, Y.E., et al., 1997, ApJ, 485, 895-868

\bibitem[\protect\astroncite{Syratovskii and Shmeleva}{1972}]{syrat72}
Syratovskii,S.I., Shmeleva, O.P., 1972, Sov. Astron., 273-285

\bibitem[\protect\astroncite{Tamres et al.}{1986}]{tamres86}
Tamres, D.H., Canfield, R.C., McClymont, A.N., 1986, ApJ, 309, 409-420

\bibitem[\protect\astroncite{Vernazza et al.}{1981}]{vernazza81}
Vernazza, J.~E., Avrett, E.~H. and Loeser, R., 1981, ApJS, 45, 635-725

\bibitem[\protect\astroncite{Vilmer et al.}{1994}]{vilmer94}
Vilmer, N., Trottet, G., Barat, C., Dezalay, J. P., Talon, R., Sunyaev, R., Terekhov, O., Kuznetsov, A., 1994, Space Science Reviews, 68, 233-238

\end{thebibliography}
\end{document}